\documentstyle[12pt]{article}
\begin{document}
\title{The $0^{++}$ and $0^{-+}$ mass of light-quark hybrid in QCD sum rules}
\author {{\small Tao Huang$^{1,2}$, Hongying Jin$^{2,3}$ and Ailin
Zhang$^2$}\\
{\small $^1$ CCAST (World Laboratory), P. O. Box 8730, Beijing, 100080}\\ 
{\small $^2$ Institute of High Energy Physics, P. O. Box 918, Beijing, 100039, 
P. R.China}\\
{\small $^3$ Fermi National Accelerator Laboratory, P. O. Box 500, Batavia,
 Illinois 60510}\\}
\date {}
\maketitle
\begin{center}
\begin{abstract}
We calculate masses of the light-quark hybrid mesons
with the quantum number $0^{++}$ and $0^{-+}$ by using the QCD sum rules.
Two kinds of the interpolated currents with the same quantum number are
employed. We find that the approximately equal mass is predicted for the 
$0^{-+}$ 
hybrid state
from the different current and the different mass is obtained for the
$0^{++}$ hybrid state from the different current. The prediction depends on
the interaction between the gluon and quarks in the low-lying hybrid mesons.
The mixing effect on the mass of the light-quark hybrid meson through Low-energy
theorem has been examined too, and  it is found that this mixing 
shifts the mass of hybrid meson and glueball a little. 
\end{abstract}
\end{center}
\section{Introduction}
\indent
\par One believes that the gluon degrees of freedom play an important role
in hadrons. QCD theory predicts the existence of glueball and hybrid states.
Searching for glueballs and hybrid mesons on experiments has been carried
for a long time since 1980s, so far there is no conclusive evidence on
them\cite{bnl}.
Glueballs and hybrids are particularly difficult to identify on experiments
since their mass spectroscopy overlaps with the ordinary $\bar qq$ meson
spectroscopy and they can mix with each other. From theoretical point of
view, glueballs and hybrids have been discussed by the bag
model \cite{bcv}, the
potential model  \cite{ct},
the flux-tube model  \cite{cp} and the QCD sum rules \cite{bdy}
\cite{lnpt}  \cite{gvgw}, but we have no effective
non-perturbative theory in QCD to predict their
mass precisely.  

\par The calculation of hybrid mesons using QCD sum rules
\cite{svz} was first given
by I. I. Balitsky {\it et al} \cite{bdy}, then another two groups J. I.
Latorre {\it et al} \cite{lnpt} and J. Govaerts {\it et al} \cite{gvgw} both
gave a revised calculation about hybrid mesons independently. There are some
errors in
their previous calculation, but they corrected them later. To avoid dealing with
the mixing of ordinary mesons with hybrids, the first two groups focused their
attention only on the hybrids with exotic quantum
numbers($J^{pc}=1^{-+},0^{--}$). They obtained the masses and decay amplitudes of
these hybrids. The third group gave the mass calculation not only
for the $J^{pc}=1^{-+},0^{--}$ hybrids but also the $J^{pc}=1^{+-},0^{++}$
hybrids. The decay width for some
decay modes of the $1^{-+}$ hybrid \cite{vg} were calculated by the last two
groups too.
\par For heavy-quark hybrids, J. Govaerts {\it et al} presented the mass calculation with
various $J^{pc}$. They analyzed sets of coupled sum rules by using the
different interpolated currents and have found that the
mass predictions for the same $J^{pc}$ from totally different sum rules
essentially agree with each other within the errors of their procedure.
These states with the same $J^{pc}$ were considered as the same states. For
light-quark hybrids, all these three groups' predictions for the mass of the
exotic $1^{-+}$ hybrid agree with the recent experiment result
\cite{bnl}. For the normal hybrids($J^{pc}=1^{-+},0^{--}$), they did not
consider the mixing effect of hybrids with ordinary mesons which is supposed
to exist. All the
calculation results from only the vector current
$g\bar q\gamma_{\alpha}G_{\alpha\mu}^aT^aq(x)$.
\par In this paper, we firstly extend the approach of J. Govaerts {\it et al} to the
light-quark case for
$0^{++}$ and $0^{-+}$ hybrids by using two kinds of the interpolated
currents:
$g\bar q\sigma_{\mu\nu}G_{\nu\mu}^aT^aq(x)$ and 
$g\bar q\gamma_{\alpha}G_{\alpha\mu}^aT^aq(x)$. 
The $\bar qq$ combination in the current  
$g\bar q\sigma_{\mu\nu}G_{\nu\mu}^aT^aq(x)$  
can be considered with the quantum number $J^{pc}=1^{+-}$ and the $\bar qq$ combination
in the current $g\bar q\gamma_{\alpha}G_{\alpha\mu}^aT^aq(x)$ has
$J^{pc}=1^{--}$, 
the interaction between quarks and gluon in these two different currents is 
different correspondingly. Thus one can't expect the same mass prediction from 
these two different
currents in the light-quark hybrid mesons sum rules. It is similar to the 
situation of
hybrid mesons in the MIT bag model \cite{bcv}. For instance, the $\bar qq$ 
combination
of $0^{++}$ hybrid meson $\bar qqg$ may have $J^{pc}=1^{--}$ with the gluon
in TE($1^{--}$) mode \cite{vw} or $J^{pc}=1^{+-}$ with the gluon in
TM($1^{+-}$) mode. These two $0^{++}$ states have different intrinsic
structure and energy. Therefore the hybrid mesons with the same $J^{pc}$ can
be obtained from totally different sum rules by using different
interpolated currents. We calculate the masses of the light-quark hybrid
mesons, $0^{++}$ and  $0^{-+}$ states, by using two different currents. Our
result shows that the prediction depends on the interaction between the
quarks and gluon in the low-lying hybrid meson. The approximately equal mass
is predicted for the $0^{-+}$ hybrid mesons from the different currents and
the different masses are obtained for the $0^{++}$ hybrid mesons from two
different currents.

\par Secondly, we consider the mixing effect on the mass determination of hybrid
meson between the
low-lying $0^{++}$ glueball and hybrid meson $\bar qqg$ . By using the
low-energy theorem \cite{al}, we can construct a sum rule for the mixing 
correlation function(one gluonic current and one hybrid current). Through
these relationship
and based on the assumption of two states (lowest-lying states
of glueball and hybrid meson $\bar qqg$) dominance,
we find the mass for the $0^{++}$ glueball
is around $1.8$ GeV, which is a little higher than the
pure resonance prediction and the mass
for the $0^{++}$ $\bar qqg$ hybrid meson is around $2.6$ GeV, which is a 
little higher than the
pure resonance prediction too.
 
\par The paper is organized as follows.  
The  analytic formalism of QCD sum rules for the hybrid is given In Sec. 2.
In Sec. 3 we give the numerical results for the mass of $0^{++}$ and
$0^{-+}$ light-quark hybrid mesons and compare them with those in the 
bag model with
the same $J^{pc}$. The mixing effect of the glueball with the hybrid meson
state is studied in Sec. 4.
The   last section is reserved for the summary. 

\section{QCD sum rules for light-quark hybrid mesons}
\indent
\par To construct the sum rules for light-quark hybrid mesons $\bar qqg$  ,
 we use the composite
operators with the same quantum numbers as these states to build the
correlation functions. In order to obtain
the $0^{++}$ hybrid meson sum rules, we define two different currents 
\begin{eqnarray}\label{current}
j(x)&=&g\bar q \sigma_{\mu\alpha}G_{\alpha\mu}^aT^aq (x) ,\\
j_{\mu}(x)&=&g\bar q \gamma_\alpha G_{\alpha\mu}^aT^aq (x) ,\label{current2} 
\end{eqnarray}
where $q (x)$ and $G_{\alpha\mu}^a(x)$ are the light-quark field and 
gluon field strength tensor, respectively.  $T^a$ are the color matrices.
\par  Through the OPE, we expand the correlation function of $j(x)$ in the
background field gauge \cite{ns} only in the  
leading order, which includes the perturbative part(a), the two-quark
condensate(b), the two-gluon condensate(c) and the four-quark condensate(d). 
The
result can be obtained from Feynman diagrams in Figs. (1a-1d)
\begin{eqnarray}\label{tensor}
\Pi(q^2)&=&{\it i}\int e^{{\it i}qx}\langle0|T\{j(x),j(0)\}|0\rangle d
x\\\nonumber
&=&-A(q^2)^3ln(-q^2/\Lambda^2)-Bq^2ln(-q^2/\Lambda^2)-Cln(-q^2/\Lambda^2)-
D\frac{1}{q^2}+const
\end{eqnarray}
where
\begin{eqnarray*}
A=\frac{\alpha_s}{24\pi^3} & , &B=\frac{4\alpha_s}{\pi}\langle m\bar qq
\rangle ,\\
C=-\frac{m^2}{\pi}\langle \alpha_s G^2 \rangle & , &D=\frac{8\pi\alpha_s}{3}
\langle m\bar qq\rangle^2 .
\end{eqnarray*}
when the u and d quarks are taken to be massless, the coefficient C vanishes.
\par In order to relate the calculation on QCD  with the hadron physics,
the standard dispersion relation is used
\begin{equation}
\Pi(q^2)=\frac{1}{\pi}
\int\frac{Im\Pi(s)}{s-q^2}ds,   
\end{equation}
the spectral density $Im\Pi(s)$ is saturated by one narrow
resonance and a continuum in a
$\theta$-function form and it is given by the following expression
\begin{equation}
Im\Pi(s)=\pi g_R^2(m_R^2)^4\delta(s-m_R^2)+\pi(As^3+Bs+C)\theta(s-s_0),
\end{equation}
where $g_R$ is the coupling of the current to the hybrid meson state and $m_R$
refers to mass of the hybrid meson.
\par In practice, it is more convenient to define the moments $R_k$ \cite{bs} 
to proceed
the sum rule calculation instead, 
which is expressed by 
\begin{eqnarray}\label{moment}
R_k(\tau, s_0)&=&\frac{1}{\tau}\hat{L}[(q^2)^k\{\Pi(Q^2)-\Pi(0)\}]
-\frac{1}{\pi}\int_{s_0}^{+\infty}s^k e^{-s\tau}Im\Pi^{\{pert\}}(s)d
s\\\nonumber
&=&\frac{1}{\pi}\int_{0}^{s_0}s^k e^{-s\tau}Im\Pi(s)d s,  
\end{eqnarray}
where $\hat{L}$ is the Borel transformation and $\tau$ is the Borel
transformation parameter, $s_0$ is the starting point of
the continuum threshold.
\par Substituting Eq.~(\ref{tensor}) into Eq.~(\ref{moment}), the 
$R_0(\tau,s_0)$ behaves as
\begin{equation}
R_0(\tau,s_0)=\frac{1}{\tau^4}\{6A[1-\rho_3(s_0\tau)]
+B\tau^2[1-\rho_1(s_0\tau)]+C\tau^3[1-\rho_0(s_0\tau)]+D\tau^4\} ,
\end{equation}
where
\begin{equation}
\rho_k(x)=e^{-x}\sum\limits_{j=0}^{k}\frac{x^j}{j!}
\end{equation}
and higher moments $R_k$ can be related to the $R_0$
\begin{equation}
R_k(\tau,s_0)=(-\frac{\partial}{\partial \tau})^kR_0(\tau,s_0).
\end{equation}
\par Similar to Eq.~(\ref{tensor}), we can calculate the correlator
$\Pi_{\mu\nu}(q^2)$ from the current $j_{\mu}(x)$ 
\begin{eqnarray}
\Pi_{\mu\nu}(q^2)&=&{\it i}\int e^{{\it
i}qx}\langle0|T\{j_{\mu}(x),j_{\nu}(0)\}|0\rangle d x\\\nonumber
&=&(\frac{q_\mu q_\nu}{q^2}-g_{\mu \nu})\Pi_v(q^2)+\frac{q_\mu
q_\nu}{q^2}\Pi_s(q^2) ,
\end{eqnarray}
and
\begin{equation}\label{pis}
\Pi_s(q^2)=-A'(q^2)^3ln(-q^2/\Lambda^2)-B'q^2ln(-q^2/\Lambda^2)-
C'ln(-q^2/\Lambda^2)-D'\frac{1}{q^2}+const,
\end{equation}
where
\begin{eqnarray*}
A'=\frac{\alpha_s}{480\pi^3} & , &B'=-(\frac{\alpha_s}{3\pi}\langle m\bar qq
\rangle+\frac{\langle\alpha_sG^2\rangle}{24\pi}) ,\\
C'=-\frac{m^2}{8\pi}\langle \alpha_s G^2 \rangle & , &D'=-\frac{2\pi\alpha_s}{3}
\langle m\bar qq\rangle^2 ,
\end{eqnarray*}
the coefficient of the two-quark condensate in $B'$ is a little different with 
reference \cite{gvgw}. The
$\Pi_v(q^2)$ is the same as this reference.
\par Replacing the $G_{\alpha\mu}^a(x)$ in Eq. ~(\ref{current}) and Eq.
~(\ref{current2}) by
\begin{equation}
\tilde 
G_{\alpha\mu}^a(x)=\frac{1}{2}\epsilon_{\alpha\mu\rho\sigma}G_{\rho\sigma}^a(x)
, 
\end{equation}
we can get the sum rules for the resonance states with opposite parity
($0^{-+}$), 
the results of the correlation functions and moments are almost the same  
as before  except that  the sign of the gluon condensate is changed.

\section{Numerical results and $J^{pc}$ analysis}
\indent
\par From Eq.~(\ref{moment}), the mass of the hybrid meson is given by 
(with $k\geq 0$)
\begin{equation}
m_R^2=\frac{R_{k+1}}{R_k},
\end{equation}
the moments $\frac{R_1}{R_0}$ and $\frac{R_2}{R_1}$ are both suitable
for the mass determination according to the ordinary QCD sum rules criteria. 
They are
employed in the following calculation.
\par To get the numerical results, 
the parameters are chosen as 
\begin{eqnarray*}
\Lambda=0.2 GeV &,&m_s=0.15 GeV ,\\
\langle\bar qq\rangle=-(0.25GeV)^3&,&\langle m\bar qq\rangle=-(0.1GeV)^4,\\
\langle m\bar ss\rangle=-0.15*0.8*0.25^3 GeV^4&,&\frac{\langle\alpha_sG^2
\rangle}{\pi}
=0.33^4 GeV^4,\\
\alpha_s(\tau)=-\frac{4\pi}{9ln(\tau \Lambda^2)}&,&
\end{eqnarray*}
where q refers to u or d quark field.
\par Corresponding to the current $j(x)$ in Eq.~(\ref{current}), 
the mass of 
$0^{++}$ $\bar ssg$ hybrid meson, determined from $\frac{R_1}{R_0}$, is shown 
as Fig. 2 and it reads $2.35$
GeV. If we use  $\frac{R_2}{R_1}$, the result is
almost the same $\sim$ $2.30$ GeV. When the quark mass vanishes, which
corresponds to $q=u, d$, the result changes a little.
\par Corresponding to the current $j_{\mu}(x)$ in Eq.~(\ref{current2}), 
$\frac{R_1}{R_0}$  gives the
mass of  $0^{++}$ $\bar ssg$ hybrid meson around $3.4$
GeV (corresponding to the dotted line in Fig. 3) and the higher moment shifts
the mass a little lower. When the quark mass goes to zero, the mass shifts a
little compared to the strange quark case.
\par There is no platform in the case of the $0^{-+}$ hybrid meson, we deal
with it as  
reference \cite{gvgw}. The masses of the $0^{-+}$ hybrid mesons 
corresponding to currents $j(x)$ and $j_{\mu}(x)$ have an approximately equal
value: $2.3$ GeV and they are shown in Fig.
4.
\par All of these results are obtained at suitable $s_0$, which account for
the ordinary QCD sum rules criteria for threshold choosing. The $s_0$ for
the current $j(x)$ is chosen as $8.0$ GeV$^2$ while the $s_0$ for the
current $j_{\mu}(x)$ is chosen as $13.0$ GeV$^2$. The results
change slightly with the $s_0$. 
\par It is apparent that the mass of the light-quark
hybrid meson depends on what interpolated current we choose:
 the mass of the $0^{++}$ hybrid from current $j_{\mu}(x)$ is about $1.0$ GeV 
higher than that from current $j(x)$ while the mass of the $0^{-+}$
hybrid from the 
two different currents is approximately the same. 
\par  We know that the hybrid meson is a three body system and the
valence quark, anti-quark and gluon may have different internal $J^{pc}$
combination.  $J^{pc}$ of $\bar q q$ shows what kind of the interaction 
between $\bar q q$  and gluon. In fact, the interaction between
quarks and gluon in the current $j(x)$ is in magnetic form, so we 
can think the $J^{pc}$ of the combination $\bar q q$ in the  current 
$j(x)$ is mainly $1^{+-}$,  while the 
interaction in the current $j_{\mu}(x)$ is in electric form and the  
$J^{pc}$ of $\bar q q$ is $1^{--}$. Only the state
with the same overall and `local' quantum number can dominate the
corresponding  correlation function, where we refer the quantum number of
intrinsic  $\bar qq$ combination or gluon, such as $J^{pc}$ of them, 
to `local' quantum number. Therefore the
correlation function which consists of the current $j(x)$ is dominated by
the $0^{++}$ state with the gluon in TE($1^{+-}$) mode and the correlation
function which consists of the current $j_{\mu}(x)$ is dominated by the
$0^{++}$ state with the gluon in TM($1^{--}$) mode. The predicted $0^{++}$
mass from the different current correspondingly  is different.  
Let's look into the diagram (c) in Fig. 1, the
$\bar\psi\sigma^{\mu\nu}\psi$ breaks helicity conservation, 
so in the center mass frame (also $\bar q-q$'s center mass frame),
 the total spin of
quark-antiquark is zero if 
quarks are massless.  On the other hand, the magnetic interaction
dominates in the current $j(x)$. Therefore, the two gluonic condensate term 
in the correlation function of $j(x)$ is
proportional to the quark's mass(it is not so for the current 
$j_\mu(x)$, see  Eq.~(\ref{tensor}) and Eq.~(\ref{pis}))
When $m$ goes to zero, this term makes a large gap 
between the masses predicted from $j(x)$ and $j_\mu(x)$.
However, in the heavy hybrid system, we have   
$m^2\sim Q^2$. This is the reason that the author in  \cite{gvgw} got the
almost same masses from two different sum rules for the heavy hybrid mesons.

\par It is helpful to note that the $J^{pc}$ of the $\bar qq$ combination in
the bag model has 
the same structure as that in the current $j_{\mu}(x)$ and $j(x)$, thus
we can compare our picture with that in the bag model. In the bag model,
the energy of the hybrid meson consist of the volume
energy, the zero-point energy, the mode energy and the $O(\alpha_s)$ quantum
corrections. The valence quarks and gluon in the $\bar qqg$ hybrid
mesons may have different excited mode and each mode has different energy.
Besides, the $O(\alpha_s)$ quantum corrections are spin dependent, and they have
different energy corresponding to different internal $J^{pc}$ combination of
the $\bar qq$ with the gluon. The
quarks and gluon in $0^{++}$ hybrid mesons may be in
$s_{\frac{1}{2}}s_{\frac{1}{2}}TM$ mode with internal 
$J^{pc}$: $1^{--}\otimes 1^{--}$ or in
$s_{\frac{1}{2}}p_{\frac{1}{2}}TE$ mode with internal 
$J^{pc}$: $1^{+-}\otimes 1^{+-}$, so the same overall $J^{pc}$ states
may  have different energy.
\section{Mixing of the $0^{++}$ hybrid meson with the glueball}
\indent 
\par In this section, we discuss the mixing effect \cite{hjz} on the mass of 
the $0^{++}$ hybrid
meson. Since the mass of the $0^{++}$ hybrid($3.4$ GeV) 
from current $j_{\mu}$ is
much larger than the pure $0^{++}$ glueball($1.7$ GeV) in sum rules, we
do not discuss this situation. Only the mixing
between the $0^{++}$ hybrid($2.3$ GeV) from current $j(x)$ and the $0^{++}$
glueball is considered. We choose the scalar gluonic current
\begin{equation}\label{glueball}
j_1(x)=\alpha_sG_{\mu\nu}^aG_{\mu\nu}^a(x)               
\end{equation}
for the $0^{++}$ glueball 
and the current
\begin{equation}\label{hybrid}
j_2(x)=g\bar q \sigma_{\mu\alpha}G_{\alpha\mu}^aT^aq (x) 
\end{equation}
for the $0^{++}$ hybrid meson.
\par The correlation function of the current in Eq.~(\ref{glueball})
 was given in  \cite{bs}
\begin{eqnarray}\label{pi}
\Pi_1(q^2)&=&a_0(Q^2)^2\ln(Q^2/\nu^2)+b_0\langle\alpha_sG^2\rangle\\\nonumber
&+&c_0\frac{\langle g
G^3\rangle}{Q^2}+d_0\frac{\langle\alpha_s^2G^4\rangle}{(Q^2)^2} ,
\end{eqnarray}
where $Q^2=-q^2>0$, and
\[
\begin{array}{rll}
a_0&=&-2(\frac {\alpha_s}{\pi})^2(1+\frac {51}{4}\frac
{\alpha_s}{\pi}),\\
c_0&=&8\alpha_s^2,\\
b_0&=&4\alpha_s(1+\frac {49}{12}\frac{\alpha_s}{\pi}),\\
d_0&=&8\pi\alpha_s,\\
\langle\alpha_sG^2\rangle&=&\langle\alpha_sG_{\mu\nu}^aG_{\mu\nu}^a\rangle,\\
\langle g G^3\rangle&=&\langle g f_{a b c}G_{\mu\nu}^a G_{\nu\rho}^b
G_{\rho\mu}^c\rangle,\\
\langle\alpha_s^2G^4\rangle&=&14\langle(\alpha_s f_{a b c}
G_{\mu\rho}^a G_{\rho\nu}^b)^2\rangle-\langle(\alpha_s f_{a b c} 
G_{\mu\rho}^a G_{\lambda\nu}^b)^2\rangle .
\end{array}
\]
\par From (\ref{moment}),(\ref{pi}) and (\ref{tensor})  
,we have the following expressions:
\begin{eqnarray}
\begin{array}{lll}
R_{0}(\tau,s_0)&=&-\frac{2a_0}{\tau^3}\lfloor1-\rho_2(s_0\tau)\rfloor
+c_0\langle gG^3\rangle+d_0\langle \alpha_s^2G^4\rangle\tau,\\
R_{1}(\tau,s_0)&=&-\frac{6a_0}{\tau^4}\lfloor1-\rho_3(s_0\tau)\rfloor
-d_0\langle\alpha_s^2G^4\rangle,\\
R_{2}(\tau,s_0)&=&-\frac{24a_0}{\tau^5}\lfloor1-\rho_4(s_0\tau)\rfloor,\\
R_0'(\tau,s_0)&=&\frac{1}{\tau^4}\{6A[1-\rho_3(s_0\tau)]
+B\tau^2[1-\rho_1(s_0\tau)]+C\tau^3[1-\rho_0(s_0\tau)]+D\tau^4\} ,\\
R_{1}'(\tau,s_0)&=&\frac{1}{\tau^5}\{24A[1-\rho_4(s_0\tau)]
+2B\tau^2[1-\rho_2(s_0\tau)]+C\tau^3[1-\rho_1(s_0\tau)]\},\\ 
\label{r}
\end{array}
\end{eqnarray}
where
$R_k$ and $R'_k$ in (\ref{r}) are the moments corresponding to
current $j_1(x)$ and  $j_2(x)$ respectively.
\par By using the Low-energy theorem  \cite{al}, we can construct another
correlator with $j_1(x)$ and $j_2(x)$ 
\begin{equation}\label{low}
\lim_{q\rightarrow 0}{\it i}\int dxe^{{\it
i}qx}\langle 0|
T[g\bar q \sigma_{\mu\alpha}G_{\alpha\mu}^aT^aq (x),\alpha_s G^2(0)] |0\rangle=
\frac{40\pi}{9}\langle 0|g\bar q\sigma_{\mu\alpha}G_{\alpha\mu}^aT^aq
|0\rangle.
\end{equation}

For the light quark,
$\langle 0|g\bar q\sigma_{\mu\alpha}G_{\alpha\mu}^aT^aq|0\rangle $ can be
expressed in terms of $\langle 0|\bar qq |0\rangle$ as \cite{rry}
\begin{equation}
\langle 0|g\bar q\sigma_{\mu\alpha}G_{\alpha\mu}^aT^aq|0\rangle
=-m^2_0\langle 0|\bar qq |0\rangle , 
\end{equation}
where $m^2_0\approx 0.8$ GeV$^2$.
\par In order to factorize the spectral density, the couplings of the
currents to the physical states are defined in the following way
\begin{eqnarray}
\langle 0|j_1|H\rangle=f_{12}m_2&,&\langle
0|j_1|G\rangle=f_{11}m_1,\\\nonumber
\langle 0|j_2|H\rangle=f_{22}m_2&,&\langle 0|j_2|G\rangle=f_{21}m_1,
\end{eqnarray}
where $m_1$ and $m_2$ refer to the glueball(including few part of quark
component) mass and the $\bar qqg$ hybrid meson(including few part of pure 
gluon component) mass, $|H\rangle$ and $|G\rangle$  refer
to the $\bar qqg$ hybrid meson state and  the glueball state respectively.
After choosing the two resonances plus continuum state approximation, the 
spectral 
density of the currents of $j_1(x)$ and $j_2(x)$ read, respectively
\begin{eqnarray}
Im\Pi_1(s)&=&m^2_2f^2_{12}\delta(s-m^2_2)+m^2_1f^2_{11}\delta(s-m^2_1)+
\frac{2}{\pi}s^2\alpha^2_s\theta(s-s_0),\\
Im\Pi_2(s)&=&m^2_2f^2_{22}\delta(s-m^2_2)+m^2_1f^2_{21}\delta(s-m^2_1)
+\pi(As^3+Bs+C)\theta(s-s_0) .
\end{eqnarray}

Then it is straightforward to get the moments
\begin{eqnarray}
R_0&=&\frac{1}{\pi}\{m^2_2e^{-m^2_2\tau} f^2_{12}+m^2_1e^{-m^2_1\tau} f^2_{11}
\},\\
R_1&=&\frac{1}{\pi}\{m^4_2e^{-m^2_2\tau} f^2_{12}+m^4_1e^{-m^2_1\tau} f^2_{11}\}
,\\
R_2&=&\frac{1}{\pi}\{m^6_2e^{-m^2_2\tau} f^2_{12}+m^6_1e^{-m^2_1\tau}
f^2_{11}\},\\
R_0'&=&\frac{1}{\pi}\{m^2_2e^{-m^2_2\tau} f^2_{22}+m^2_1e^{-m^2_1\tau} 
f^2_{21}\},\\
R_1'&=&\frac{1}{\pi}\{m^4_2e^{-m^2_2\tau} f^2_{22}+m^4_1e^{-m^2_1\tau} 
f^2_{21}\}.
\end{eqnarray}
In the meantime, assuming the states $|G\rangle$ and $|H\rangle$ saturate
the l.h.s of Eq. (\ref{low}), one can obtain 
\begin{equation}
\lim_{q\rightarrow 0}{\it i}\int dxe^{{\it
i}qx}\langle 0|
T[g\bar q \sigma_{\mu\alpha}G_{\alpha\mu}^aT^aq (x),\alpha_s G^2(0)] 
|0\rangle=
f_{22}f_{12}+f_{21}f_{11}.
\end{equation}
To get the numerical result, the following additional parameters are chosen
\begin{eqnarray*}
\langle gG^3\rangle&=&(0.27 GeV^2)\langle\alpha_sG^2\rangle,\\
\langle\alpha_s^2G^4\rangle&=&\frac{9}{16}\langle\alpha_sG^2\rangle^2,\\
\end{eqnarray*}

The next step is to equate the QCD side with the
hadron side one by one, and we get
a set of equations. Given various of reasonable parameters $s_0$ and 
$\tau$, a series of masses of the two states through solving this set of
equations are obtained. Our result is illustrated in Fig. 5.
The doted line
corresponds to the hybrid meson and the solid line corresponds to the
glueball in this figure. It is shown  
that $s_0=8.0$ GeV$^2$ is the best favorable value, then from the figure
follows the masses 
prediction: the hybrid meson with mass around $2.6$ GeV 
and the glueball with mass around $1.8$ GeV.
We conclude that the mixing makes the masses of the glueball and the hybrid 
meson are both a little
higher than their  pure states.
\section{Summary}
\indent
\par In this paper, we calculate the $0^{++}$ and $0^{-+}$ masses of the 
light-quark hybrid
mesons by using the QCD sum rules with two different kinds of interpolated
currents: $g\bar q\sigma_{\mu\nu}G_{\nu\mu}^aT^aq(x)$ and
$g\bar q\gamma_{\alpha}G_{\alpha\mu}^aT^aq(x)$. Numerical result shows that  
the $0^{++}$ hybrid meson mass from current $j_{\mu}(x)$ 
is around $3.4$ GeV, which is $1.0$ GeV
higher than that(which is around $2.35$ GeV) from current $j(x)$. The masses
of the $0^{-+}$ hybrid mesons
from these two currents are approximatively equal: $2.3$ GeV.
\par The $J^{pc}$ of the $\bar qq$ combination in these two current are
$1^{+-}$ and $1^{--}$ respectively, so the interaction between the quarks
and gluon is different and the two different kind of $0^{++}$
or $0^{-+}$ hybrid
meson dominating the spectral density of these two different current
are different states correspondingly. For the light-quark hybrids, the
interaction between the quarks
and gluon makes a large contribution to the energy of the states, their mass 
thus may be
different, while for heavy-quark hybrid, $m^2\sim Q^2$, so different
currents result in the approximately equal mass prediction.
Compared to MIT
bag model, our picture confirms the reasonness of the mode analysis in the bag 
model.
\par
For the $0^{++}$ hybrid, the contribution of the two-gluon
condensate to the correlation function( Eq.~(\ref{pis})) from the current
$j_{\mu}(x)$ is large, while the contribution of the two-gluon
condensate to the correlation function( Eq.~(\ref{tensor})) from the current
$j(x)$ is small because of
the factor $m^2$ in the coefficient C, these two $0^{++}$ states have 
different mass value.
For the $0^{-+}$ hybrid, the sign of two-gluon condensate terms in
Eq.~(\ref{pis}) and Eq.~(\ref{tensor}) changes. This change affects slightly
the correlation function and the mass prediction from the current $j(x)$ since 
the two-gluon
condensate contribution is small in it. However, in the case of the current
$j_\mu(x)$ the two-gluon condensate contribution to the correlation function
is large, the change of sign makes the correlation function a big
difference from the $0^{++}$ hybrid meson and it results in a much lower
mass prediction for the $0^{-+}$ hybrid.  These two  $0^{-+}$ light hybrid 
from the
two different currents thus have
approximately equal mass.
\par The mixing effect on the mass determination of $0^{++}$ hybrid meson is
considered too. We find
that the mixing of the $0^{++}$ hybrid meson with the glueball shifts
the masses of both the hybrid and the glueball a little higher than their
pure states.  
\\

{\bf Acknowledgment}

This work is supported in part by the national natural science foundation 
of P. R. China. H. Y. Jin thanks Theoretical Physics Division
at  Fermilab  for their kind hospitality.

\newpage
\par
{\huge\bf Figure caption}\\
\par
Figure 1: Feynman diagrams of the leading order contributing to the
correlation function.\\
\par
\par
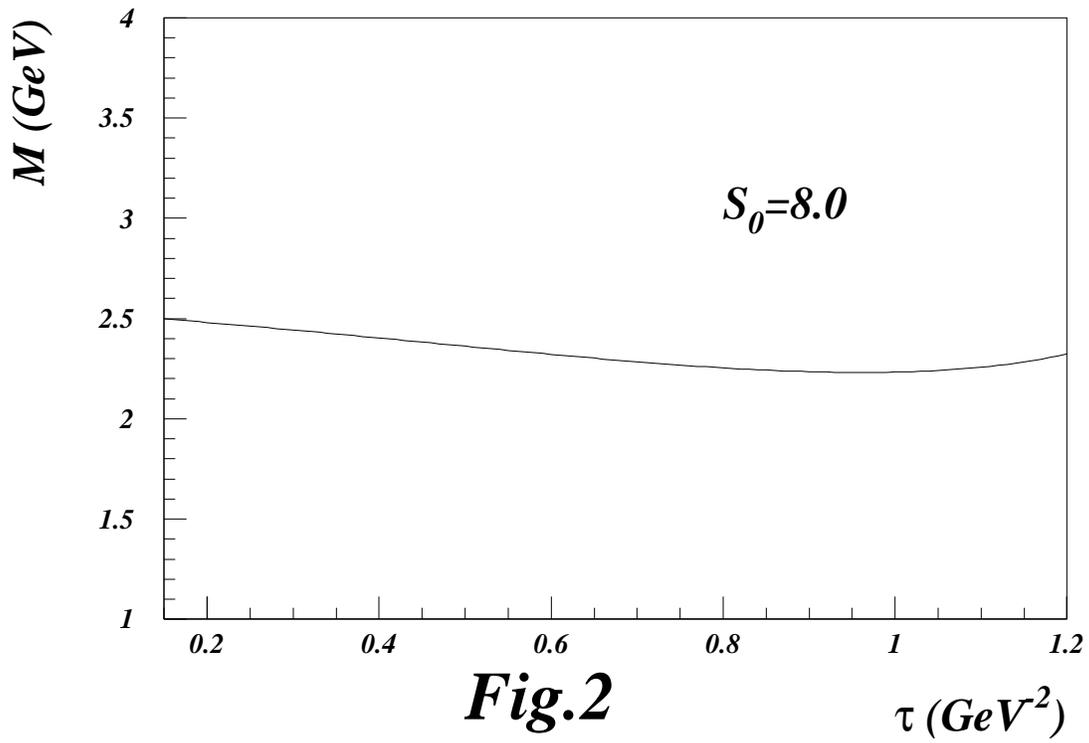
Figure 2: $0^{++}$ $\bar ssg$ mass from $\frac{R_1}{R_0}$ versus $\tau$ at 
$s_0=8.0$ GeV$^2$
corresponding to current $j(x)$.\\
\par
\par
Figure 3: $0^{++}$ hybrid mass from $\frac{R_1}{R_0}$ versus $\tau$ at 
$s_0=13.0$
GeV$^2$ corresponding to current $j_{\mu}(x)$.\\
\par
\par
Figure 4: $0^{-+}$ hybrid mass from $\frac{R_1}{R_0}$ versus $\tau$ at
$s_0=8.0$ GeV$^2$ corresponding to current $j(x)$ and $j_{\mu}(x)$.\\
\par
\par
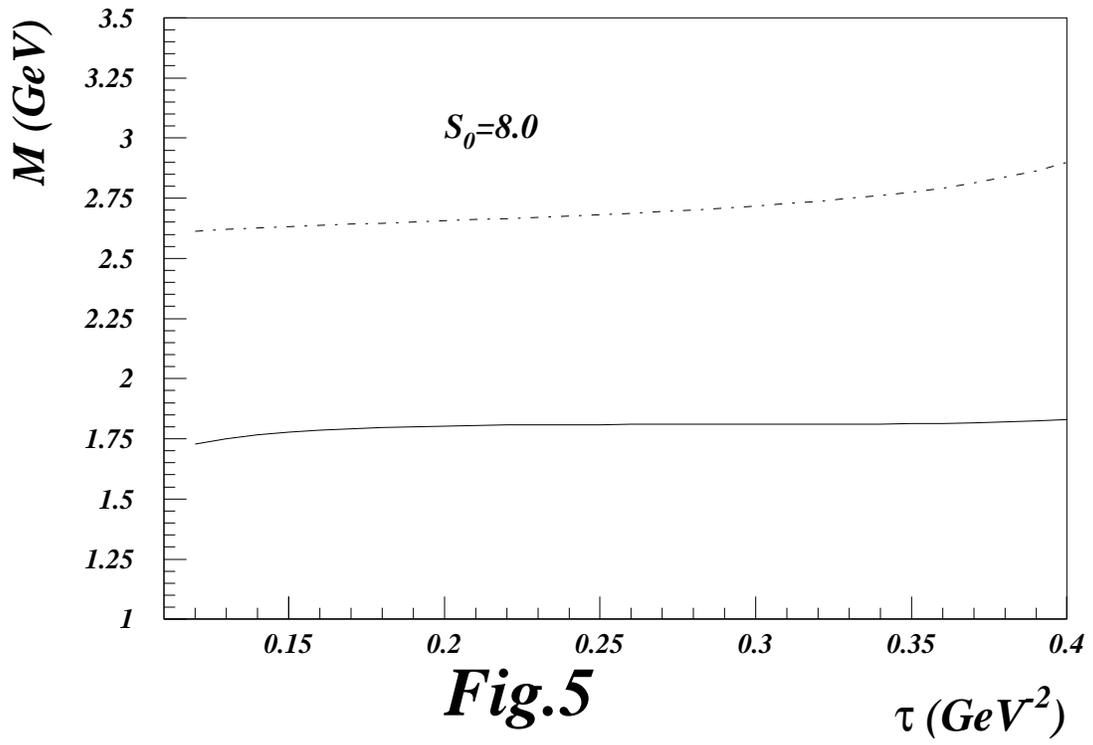
Figure 5: $0^{++}$ $\bar ssg$ mass versus $\tau$ at $s_0=8.0$ GeV$^2$
corresponding to the mixing figure.
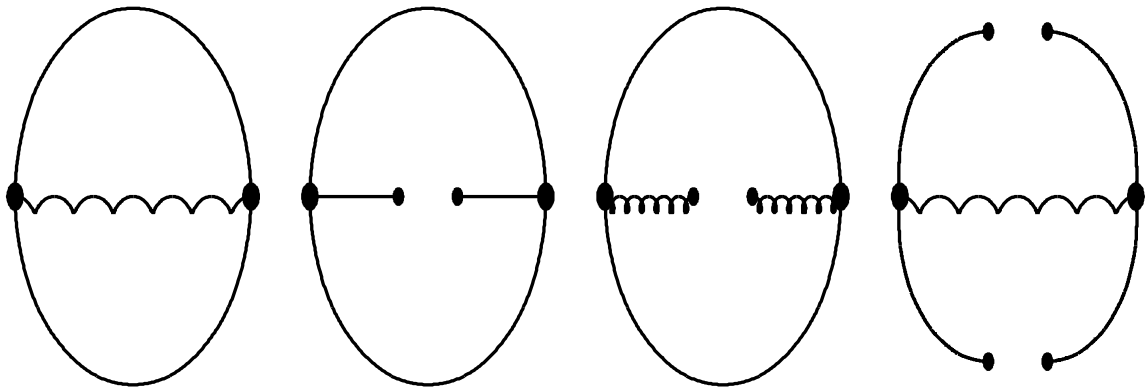
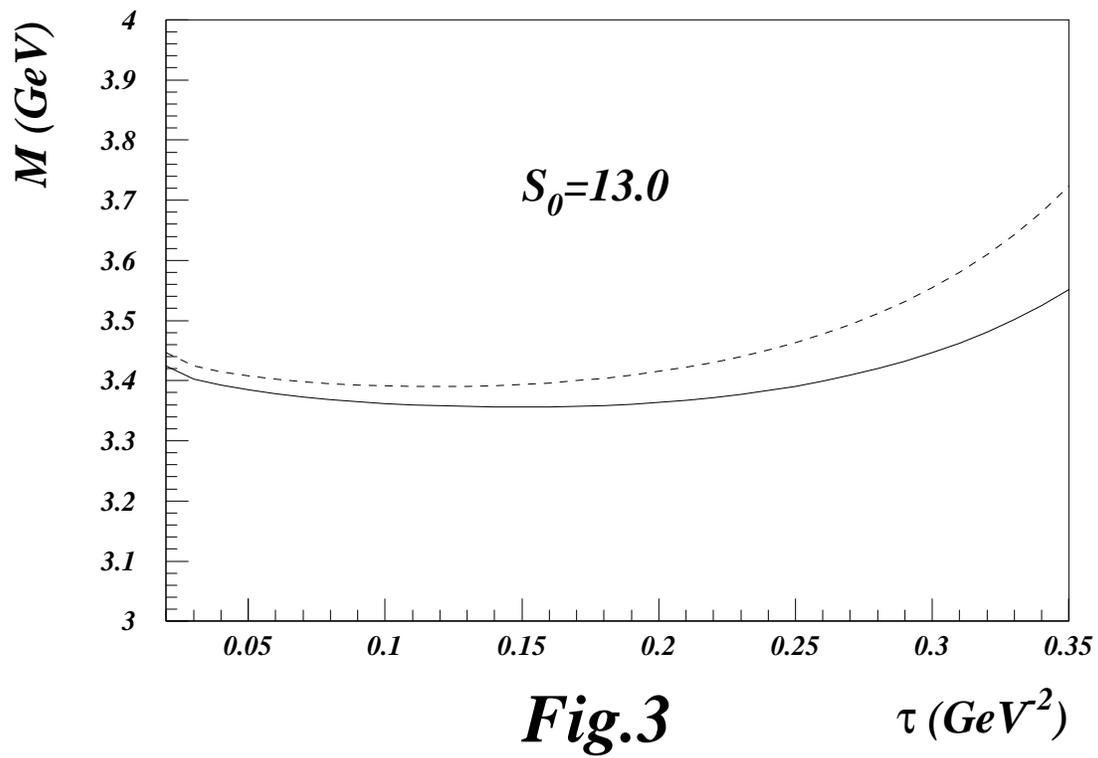
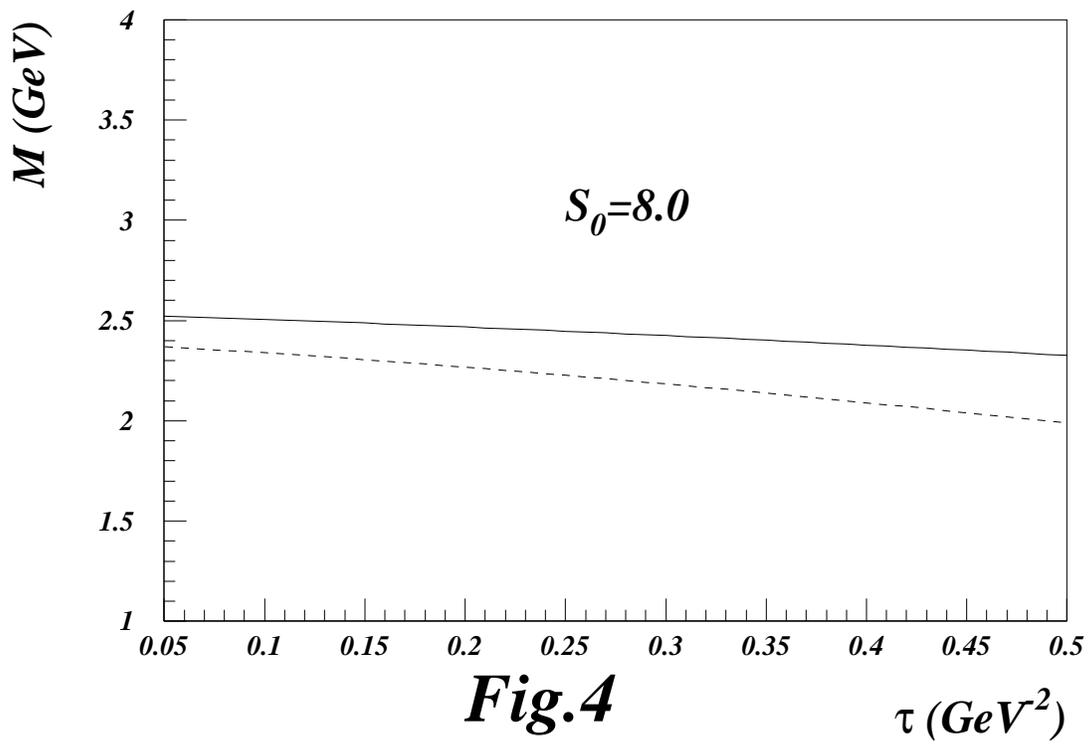
\end{document}